\documentclass[twocolumn,amsmath,amssymb,prl,superscriptaddress,floatfix]{revtex4}
\usepackage{graphicx}
\usepackage{bm}
\usepackage{amsmath,amssymb}

\begin{document}
\title{Non-equilibrium coherence dynamics in one-dimensional Bose gases}

\author{S.~Hofferberth}
\affiliation{Atominstitut der \"Osterreichischen Universit\"aten,
TU-Wien, Stadionallee 2, A-1020 Vienna, Austria }
\affiliation{Institut f\"ur Experimentalphysik, Universit\"at
Heidelberg, Philosophenweg 12, D-69120 Heidelberg, Germany}
\author{I.~Lesanovsky}
\affiliation{Institut f\"{u}r Theoretische Physik, Universit\"{a}t
Innsbruck, Technikerstr. 21a, A-6020 Innsbruck, Austria}
\author{B.~Fischer}
\affiliation{Institut f\"ur Experimentalphysik, Universit\"at
Heidelberg, Philosophenweg 12, D-69120 Heidelberg, Germany}
\author{T.~Schumm}
\affiliation{Atominstitut der \"Osterreichischen Universit\"aten,
TU-Wien, Stadionallee 2, A-1020 Vienna, Austria }
\author{J.~Schmiedmayer}
\email{schmiedmayer@atomchip.org} \affiliation{Atominstitut der
\"Osterreichischen Universit\"aten, TU-Wien, Stadionallee 2, A-1020
Vienna, Austria } \affiliation{Institut f\"ur Experimentalphysik,
Universit\"at Heidelberg, Philosophenweg 12, D-69120 Heidelberg,
Germany}

\maketitle

\textbf{Low-dimensional systems are beautiful examples of many-body
quantum physics \cite{Popov1983}. For one-dimensional systems
\cite{Giamarchi2003} the Luttinger liquid approach
\cite{Haldane1981} provides insight into universal properties. Much
is known of the equilibrium state, both in the weakly
\cite{Petrov2000,Dettmer2001,Richard2003,Pricoupenko2004} and
strongly \cite{Paredes2004,Kinoshita2004} interacting regime.
However, it remains a challenge to probe the dynamics by which this
equilibrium state is reached \cite{Kinoshita2006}. Here we present a
direct experimental study of the coherence dynamics in both isolated
and coupled degenerate 1d Bose gases. Dynamic splitting is used to
create two 1d systems in a phase coherent state \cite{Schumm2005}.
The time evolution of the coherence is revealed in local phase
shifts of the subsequently observed interference patterns.
Completely isolated 1d Bose gases are observed to exhibit a
universal sub-exponential coherence decay in excellent agreement
with recent predictions by Burkov et al. \cite{Burkov2007}. For two
coupled 1d Bose gases the coherence factor is observed to approach a
non-zero equilibrium value as predicted by a Bogoliubov approach
\cite{Whitlock2003}. This coupled-system decay to finite coherence
is the matter wave equivalent of phase locking two lasers by
injection. The non-equilibrium dynamics of superfluids plays an
important role in a wide range of physical systems, such as
superconductors, quantum-Hall systems, superfluid Helium, and spin
systems \cite{Blatter1994,Shimshoni1998,Forte1992}. Our experiments
studying coherence dynamics show that 1d Bose gases are ideally
suited for investigating this class of phenomena.}

The starting point of our experiments is a 1d Bose gas of a few
thousand atoms trapped in a highly elongated, cylindrical magnetic
microtrap on an atom chip \cite{Folman2002,Fortagh2007} with
typical transverse and longitudinal oscillator frequencies of
$\nu_\perp \sim 4.0$\,kHz  and $\nu_z \sim  5$\,Hz. Our trapped
Bose gas is in the 1d quasi-condensate regime \cite{Popov1983},
characterized by both the temperature $T$ and chemical potential
$\mu$ fulfilling $k_\text{B} T, \mu < h \nu_\perp$
\cite{Bouchoule2007}.

\begin{figure}
\includegraphics[angle=0,width=\columnwidth]{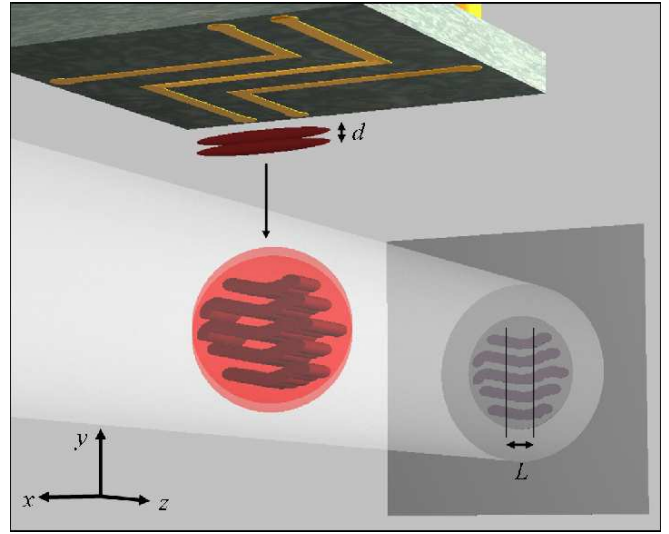}
\caption{\label{Fig:setup} Schematic of the experiment. A single
1d quasi-condensate is phase coherently split using rf potentials
on an atom chip. A combination of two rf fields allows balanced
splitting in the vertical direction \cite{Hofferberth2006}. After
the separation, the system is held in the double-well potential
for a variable time $t$ and is then released from the trap. The
resulting interference pattern is imaged along the transverse
direction of the system. Thermal phase fluctuations in the two
quasi-condensates can be directly observed as local shifts of the
observed fringe profiles.}
\end{figure}

After the initial preparation of this single 1d system, we perform
a phase-coherent splitting along the transverse direction by means
of a radio-frequency (rf) induced adiabatic potential
\cite{Schumm2005}. As shown in figure \ref{Fig:setup}, the final
system consists of two 1d quasi-condensates in a vertically
orientated double-well potential \cite{Hofferberth2006}. They are separated by a tunable potential barrier, whose height is controlled by the applied rf fields.

This splitting process initializes the system in a mutually phase
coherent state. Directly after the splitting, the phase
fluctuation patterns of the two individual quasi-condensates are
identical, resulting in a vanishing global relative phase. This is
a highly non-equilibrium state of the split system, which will
relax to equilibrium over time.

\begin{figure}
\includegraphics[angle=0,width=\columnwidth]{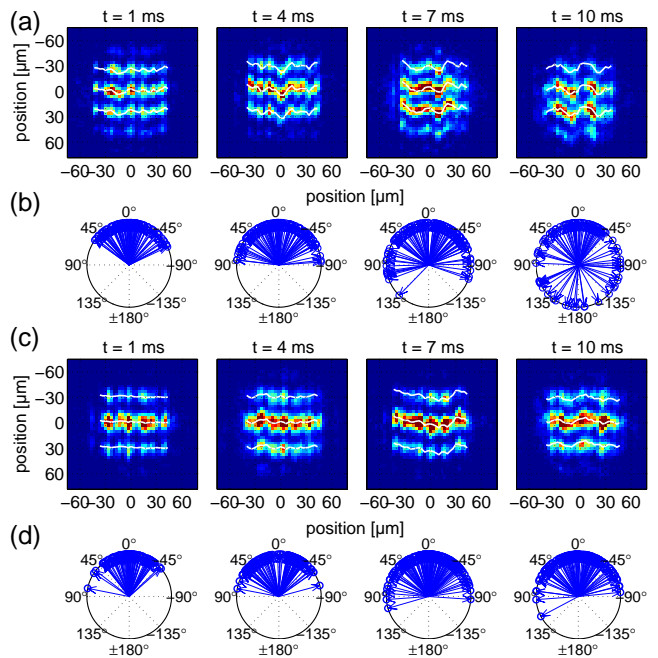}
\caption{\label{Fig:examples} Direct observation of the phase
dynamics through interference. (a) Example images of the observed
interference patterns for hold times $t=1,4,7,10$\,ms in the case
of isolated 1d systems (taken from the $T=133$ nK data set in
figure \ref{Fig:uncoupled}b). White lines indicate the bright
nodes of the local interference pattern, extracted from fits to
the profiles in each vertical pixel slice. The randomization of
the relative phase $\theta(z,t)$ over time leads to an increased
waviness of the observed fringe patterns. (b) Corresponding polar
plots of the local relative phases from the central regions of
five images. For the isolated systems we find a complete
randomization of the local relative phase over time. (c) For a
finite tunnel coupling through the potential barrier, the phase
randomization is counterbalanced by particle tunnelling, resulting
in a suppression of the fringe waviness (images correspond to
figure \ref{Fig:tunneling}(c)). (d) This results in a non-random
steady-state phase spread, whose width depends on the tunnel
coupling and the temperature of the system.}
\end{figure}

To study this time evolution of the phase coherence, the two 1d
Bose gases are held in the double-well configuration for a varying
time $t$ before they are released and recombined in
time-of-flight. The resulting interference pattern is recorded
using standard absorption imaging along the transverse direction
of the system. The spatial variation of the relative phase between
the two quasi-condensates translates directly into local phase
shifts in the interference pattern (Figure \ref{Fig:examples}).

If the two parts of the system are completely separated (compare
methods), the equilibrium state consists of two uncorrelated
quasi-condensates. Consequently, we observe an increasing waviness
of the interference pattern with time, which in the end leads to a
complete randomization of the relative phase $\theta(z,t)$ (Figure
\ref{Fig:examples}a, b). This change in the interference pattern is
a direct visualization of the dynamics of the phase fluctuations.
Qualitatively similar behavior was recently observed at MIT
\cite{Jo2007b} for elongated condensates with $\mu \sim 2 h
\nu_\perp$ and $T \sim 5 h \nu_\perp$.

For a finite tunnel coupling (compare methods) between the two
systems, we also observe an increase in the waviness of the
interference. However, in contrast to the completely separated case,
the final equilibrium state shows a non-random phase distribution
(Figure \ref{Fig:examples}c,d). This is caused by the phase
randomization being counterbalanced by the coherent particle
exchange between the two fractions. The final width of the observed
phase spread depends on the strength of the tunnel coupling
\cite{Spietz2003,Gati2006b}.

A quantitative measure of the fluctuations of the local
relative phase is given by the coherence factor
\begin{eqnarray} \label{coh_fac_def}
  \Psi(t)=\frac{1}{L} \left|\int dz\, e^{i\theta(z,t)}\right|
\end{eqnarray}
where $L$ is the length of the analyzed signal.

We obtain the coherence factor from a single image by extracting
the local relative phase $\theta(z,t)$ from the interference
pattern in each vertical pixel slice. To account for variations
in the 1d atomic density $n_\text{1d}(z)$ due to the longitudinal
confining potential, we use only the central $40\%$ of each image
in our analysis. Over this range, $n_\text{1d}(z)$ varies by only
$\sim 15\%$ from its peak value. In the following, we neglect this
modulation and assume a homogeneous density, obtained by averaging
over the density profile in this center region.

\begin{figure}
\includegraphics[angle=0,width=\columnwidth]{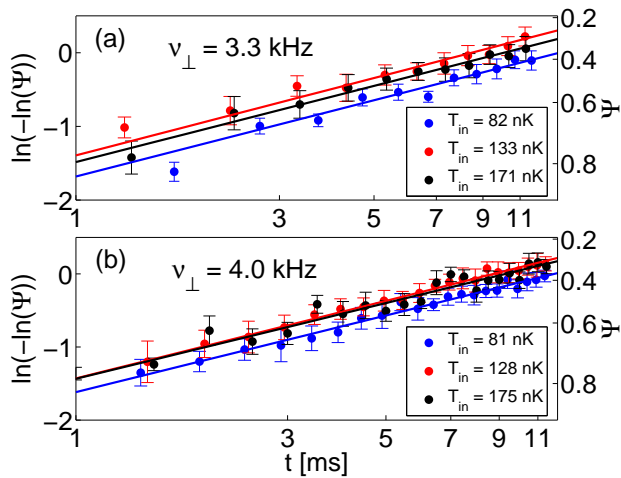}
\caption{\label{Fig:uncoupled}Time evolution of the coherence
factor for uncoupled 1d quasi-condensates. Double logarithmic
plots of $\ln(\Psi)$ as a function of time $t$ for transverse
confinements of (a) $\nu_\perp = 3.3$\,kHz and (b) $\nu_\perp =
4.0$\,kHz, respectively. Three different data sets corresponding
to different initial temperatures and line densities are shown for
each configuration. Each data point is the average of 15
individual measurements. Error bars indicate the standard error of
the mean. In this representation, the exponent of the decay
function is given directly by the slope of the observed linear
evolution of $\ln(\Psi)$. The lines shown are linear fits to the
data, the obtained slopes are $0.64 \pm 0.08$, $0.65 \pm 0.07$,
and $0.64 \pm 0.04$ for (a) and $0.65 \pm 0.03$, $0.66 \pm 0.03$,
and $0.64 \pm 0.06$ for (b). All slopes show good agreement with
the theoretical prediction of $2/3$.}
\end{figure}

We first investigate the time evolution of the coherence factor
$\Psi(t)$ for the case of completely separated 1d
quasi-condensates, separated by a potential barrier sufficiently
high to suppress any tunnel coupling (Figure
~\ref{Fig:examples}a,b). The time evolution of the measured
coherence factor for six different combinations of initial
temperature $T_\text{in}$, 1d density $n_\text{1d}$, and
transverse trap frequency $\nu_\perp$ is displayed in figure
\ref{Fig:uncoupled}. In a time window of 1-12 ms we observe a
universal sub-exponential decay of the form
\begin{eqnarray}
\label{Coherence_factor}
  \Psi(t)\propto e^{-(t/t_0)^{\alpha}},
\end{eqnarray}
with decay time constants $t_0$ ranging from 5\,ms to 9\,ms
(Table~\ref{table}).
\begin{table}
\begin{tabular}{|c|c|c|c|c|c|c|}
   \hline
   $T_\text{in}$ & $n_\text{1d}$ & $\mu/h$ & $\nu_\perp$  & $\alpha$ & $t_0$ & $T_\text{f}$\\
   $[nK]$        & [$1/\mu\text{m}$] & [kHz] & [kHz] &         &[ms]& [nK] \\
   \hline $ 82(28)$ & $20(4)$ & $0.7(1)$& $3.3$ & $0.64(8)$ &$9.0(4)$ & $76(10)$ \\
   \hline $133(25)$ & $34(5)$ & $1.2(2)$& $3.3$ & $0.65(7)$ &$5.5(3)$ & $145(13)$ \\
   \hline $171(19)$ & $52(4)$ & $1.8(1)$& $3.3$ & $0.64(4)$ &$6.4(3)$ & $186(15)$ \\
   \hline $ 81(31)$ & $22(4)$ & $0.9(2)$& $4.0$ & $0.65(3)$ &$8.1(2)$ & $85(10)$ \\
   \hline $128(23)$ & $37(4)$ & $1.5(2)$& $4.0$ & $0.66(3)$ &$5.9(2)$ & $153(13)$\\
   \hline $175(20)$ & $51(5)$ & $2.1(2)$& $4.0$ & $0.64(6)$ &$6.1(4)$ & $194(17)$ \\
   \hline
\end{tabular}
\caption{\label{table} Measured exponents $\alpha$, decay time
constants $t_0$ and final temperatures $T_{f}$ after the splitting
for the data shown in figure \ref{Fig:uncoupled}.}
\end{table}

Our results can be directly compared with a recent theoretical
study by Burkov \textit{et al.} \cite{Burkov2007} based on a
Luttinger liquid approach \cite{Haldane1981}, which predicts
$\alpha=2/3$ for the exponent in Eq.~\ref{Coherence_factor}. For all the cases studied, we find $\alpha$ to be between 0.64 and
0.67 (Table~\ref{table}), in very good agreement with the
theoretical prediction. This agreement is even more remarkable
because the approximations used in ref. \cite{Burkov2007} to
obtain Eq.~\ref{Coherence_factor} and $\alpha=2/3$ are only valid
for $t>t_0$, whereas our data covers the full range from 0 up to 2
$t_0$.

At our typical temperatures ($T_\text{in} \approx 80-150$ nK), the
dynamics of the coherence factor is dominated by thermal phase
fluctuations \cite{Bistritzer2007,Burkov2007}. Burkov \textit{et
al.} derive the decay of $\Psi(t)$ by assuming that the system of
two 1d Bose gases can evolve into thermal equilibrium. The
sub-exponential coherence decay then results from the fact that
damping in a 1d liquid at finite temperature is always
non-hydrodynamic, which in turn is caused by the break-down of
superfluid order in 1d on length scales longer than the
temperature-dependent correlation length \cite{Andreev1980}. The
experimentally observed decay agrees with this non-trivial
theoretical prediction, which is strong evidence that the final
equilibrium state of our system is truly thermal equilibrium. In
turn, this suggests the non-integrability of the experimentally
realized system \cite{Rigol2007}.

Following ref. \cite{Burkov2007} the decay time constant $t_0$
can be related to the parameters of the 1d quantum gas:
\begin{eqnarray}
\label{Eq_t0}
    t_0=2.61\,\pi g n_\text{1d} K/T_\text{f}^2
\end{eqnarray}
where $K = \pi \hbar \sqrt{\frac{n_\text{1d}}{g m}}$ is the
Luttinger parameter for the weakly interacting 1d Bose gas
\cite{Cazalilla2004}. $g = 2 h \nu_\perp a_s$ is the effective
1d coupling constant with $a_s$ being the s-wave scattering
length, $m$ the mass of the atoms, and $T_\text{f}$ the final
equilibrium temperature of the split system.

$T_\text{f}$ can be extracted from the decay constant $t_0$
following Eq.~\ref{Eq_t0} using independently measured values
of $\nu_\perp$ and $n_\text{1d}$. The
results, together with an independently measured initial
temperature of the unsplit system $T_\text{in}$ from
time-of-flight images, are compiled in table \ref{table}. Within
the error bounds, the temperatures $T_\text{in}$ and $T_\text{f}$
agree. Never the less, our data seems to indicate an increasing difference between $T_\text{in}$ and $T_\text{f}$ for increasing initial temperature.

This observation is in qualitative agreement with ref
\cite{Burkov2007}, which predicts the heating during the splitting
to scale with $\sqrt{T_\text{in} K}$. This is also supported by
the observation that for identical $T_\text{in}$, we find the same
temperature increase in the split system for both of the two
different transverse confinements used in the experiment.

One should note here that the temperature $T_\text{f}$ is the
temperature of the longitudinal excitations and is measured on a
time scale much shorter than the characteristic time given by the
sound propagation through the sample, or a thermalization
timescale, which should be even longer. To check the universality
of our experiments, we cut each sample in two and analyze each
half separately. We obtain the same results within the statistical
uncertainties.

\begin{figure}
\includegraphics[angle=0,width=\columnwidth]{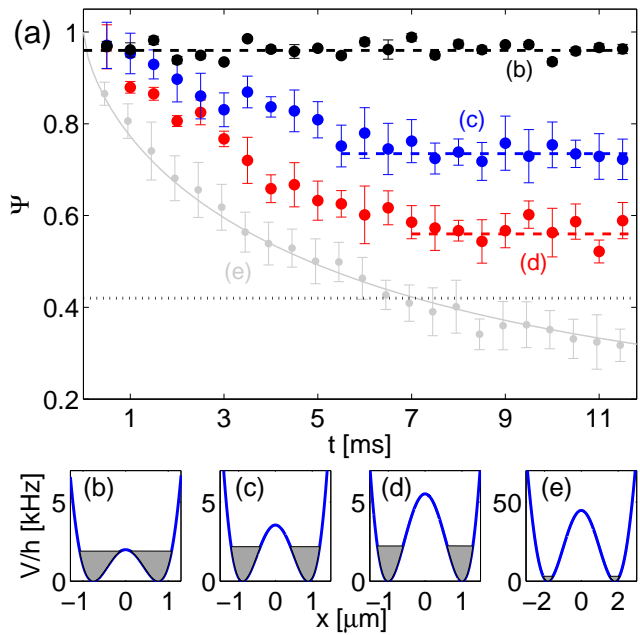}
\caption{\label{Fig:tunneling}Time evolution of the coherence
factor for coupled 1d quasi-condensates. (a) After an initial
decay we observe a time-independent coherence factor determined by
the strength of the tunnel coupling. Different colors correspond
to the different transverse double-well potentials shown in (b-d)
as indicated. Each data point is the average of 12 individual
measurements. Error bars indicate s.e.m. The light grey points are
the $T=128$ nK data set from figure \ref{Fig:uncoupled}b, showing
the sub-exponential decay in the case of uncoupled condensates for
comparison. The corresponding transverse potential is shown in
(e). The dotted black line in (a) indicates the Rayleigh test
confidence level above which the considered phase distribution of
a single measurement can be identified as non-random with a $90\%$
probability. In the coupled case, the final phase distribution in
each realization can be clearly considered as non-random.}
\end{figure}

We now turn to the case of (weakly) coupled 1d quasi-condensates
realized when the potential barrier between the two 1d systems is
only slightly larger than the chemical potential $\mu$, and
tunnelling between the two split systems is significant. In this
case we observe a qualitative change in the behavior of the
coherence factor $\Psi(t)$, as illustrated in figure
\ref{Fig:tunneling}. After an initial decay, the coherence factor
saturates well above the value for random phases. This observation
is a clear manifestation of a balancing between phase locking due
to coherent tunnelling and the local phase fluctuations in the 1d
quasi-condensate \cite{Whitlock2003}. It is the matter wave
equivalent of injection locking two lasers.

For a quantitative analysis, we use the results of ref.
\cite{Whitlock2003} for the energy spectrum and the mode functions
and express the final equilibrium coherence factor as
\begin{eqnarray}
\label{Psi_tunneling}
 \Psi = \Psi_{q} \times \exp \left(-\frac{1}{4\pi n_\text{1d}}
 \int \text{d}k \frac{S_k}
 {e^{\frac{\epsilon_k}{k_\text{B}T}}-1}\right),
\end{eqnarray}
where $\Psi_{q} (\approx 1$ in our case) is the contribution of
the quantum fluctuations,  and
\begin{eqnarray}
\epsilon_k &=& \sqrt{\left(\frac{\hbar^2 k^2}{2 m} + 2\gamma
\right)\left(\frac{\hbar^2 k^2}{2 m} + 2\gamma +2 g
n_\text{1d}\right)} \nonumber \\
S_k &=& \sqrt{\frac{\frac{\hbar^2 k^2}{2 m} + 2\gamma +2 g
n_\text{1d}}{\frac{\hbar^2 k^2}{2 m} + 2\gamma }}
\end{eqnarray}
are the energy eigenvalues and amplitudes of the Bogoliubov modes,
respectively. $\gamma$ thereby quantifies the tunnel coupling
between the two 1d systems. Independent measurements of $T$ and
$n_\text{1d}$ allow us to extract $\gamma$ directly from the
measured equilibrium coherence factors. For the red, blue, and
black data sets in figure \ref{Fig:tunneling}, we obtain
$\frac{\gamma}{g n_\text{1d}} \approx 0.003, 0.01$, and $0.2$,
respectively. Comparison with numerical calculations of the tunnel
coupling, based on the spatial overlap of the wave functions
\cite{Ananikian2005}, yields qualitative agreement within a factor
of three. This suggests that a two-mode model for the tunnelling
dynamics is not sufficient to describe our experiments.
Alternatively, if $\gamma$ can be determined independently, the
equilibrium coherence factor can be used for thermometry in the
coupled system \cite{Gati2006b}.

Particle exchange between the two condensates takes place on a
timescale given by the inverse of the Josephson oscillation
frequency \cite{Whitlock2003}
\begin{eqnarray}
\omega_\text{J}= \frac{\sqrt{ \gamma g n_\text{1d}}}{\hbar}.
\end{eqnarray}
>From our measured $\gamma$ we estimate $\omega_\text{J} \approx 2
\pi \times 80, 200, 900$ Hz for the three presented data sets,
which should be compared to the coherence decay time scale of the
order of 5 ms (Figure \ref{Fig:tunneling}). This suggests that the
balancing process between coherent tunnelling and phase
fluctuations stabilizes on timescales of the order of a single or
few Josephson oscillations in a coupled 1d system.

In summary, we have studied the non-equilibrium phase dynamics in
a 1d Bose gas with the aid of matter wave interferometry. For
completely separated systems we observe the sub-exponential
coherence decay predicted by Luttinger liquid theory
\cite{Burkov2007}. For coupled 1d Bose gases, the balance between
decoherence and phase stabilization due to tunnelling establishes
a finite coherence \cite{Whitlock2003,Gati2006b}. In both examples
our experiments illustrate how the thermal equilibrium state of a
phase-fluctuating 1d quasi-condensate
\cite{Popov1983,Petrov2000,Dettmer2001,Richard2003,Pricoupenko2004}
is approached from the highly non-equilibrium situation of two
phase coherent copies created by splitting a degenerate 1d Bose
gas.

The above investigation demonstrates that with the exquisite
control provided by atom chips, one can now study in detail the
non-equilibrium dynamics of coherence in a single quantum
degenerate 1d system. This allows us to address the fundamental
issue of thermalization in (almost) integrable systems
\cite{Rigol2007}. Our approach can easily be extended to study the
dynamics of two- and three-dimensional quantum systems, and to
include other factors such as internal state dynamics. Moreover,
it will be the starting point for detailed studies of coupled 1d
quantum systems. One prominent example is the quantum Sine-Gordon
model, which plays an important role in many fundamental physics
questions
\cite{Giamarchi2003,Bouchoule2005,Gritsev2007,Gritsev2007b}.

\subsection{Acknowledgement}
We thank A Burkov, V. Gritsev, E. Demler, R. Bistritzer, and E.
Altman for stimulating discussions. We also thank S. Groth for
fabricating the atom chip used in the experiments. We acknowledge
financial support from the Wittgenstein Prize and the European
Union, through Atom Chips and FET/QIPC SCALA projects.

\section*{Methods}

\subsection{Preparing a 1d BEC on an atom chip}
We start by preparing a magnetically trapped ultra cold ($T \sim 1
\mu$K) sample of Rb$^{87}$ atoms in the $|F=2,m_\text{F}=2>$ state
on the atom chip \cite{Folman2002,Fortagh2007} using our standard
procedure of laser cooling, magnetic trapping and evaporative
cooling \cite{Wildermuth2004}. The sample is then transferred to a
highly elongated magnetic trap ($\nu_\perp \sim 4.0$\,kHz,
longitudinal confinement $\nu_z < 5$\,Hz) at a distance of $75
\,\mu$m from the atom chip surface and cooled to quantum
degeneracy by tuning the evaporative cooling radio frequency (rf)
in between the ground and first excited transverse trapping state.
The resulting sample an effectively one-dimensional
\cite{Petrov2000,Bouchoule2007} with $\mu, k_\text{B} T < h
\nu_\perp$ containing $1...10 \times 10^3$ atoms.

\subsection{Smoothness of the trapping potentials}
Our atom chip wires \cite{Groth2004} provide exceptionally smooth
trapping potentials \cite{Krueger2004c}.   The residual roughness
due to current-flow perturbations in the trapping wire is much too
small to be measured directly at our operation distance $h = 75
\,\mu$m from the atom chip surface. An estimate based on the
current flow pattern in the wire reconstructed from magnetic field
microscopy measurements employing 1d condensates at short distance
($\sim 10 \,\mu$m) \cite{Wildermuth2006} shows that for length
scales below $40 \,\mu$m, the remaining potential perturbations
are smaller than $1 \,\mu$G (corresponding to $<10^{-3} \mu$). Due
to the exponential damping of the higher spatial modes (wave
vector $k=\frac{2 \pi}{\lambda}$) of the magnetic field variations
with distance $h$ from the chip surface following $exp(-k h)$, the
disorder potentials will be even smaller on the short length
scales corresponding to the phase coherence length or the healing
length in the 1d Bose gas.

\subsection{Preparing two phase-coherent 1d systems}
To prepare two mutually phase coherent 1d systems, we employ a
combination of static and rf magnetic fields on our atom chip.
Coupling electronic ground states of the magnetically trapped
atoms results in dressed-state adiabatic potentials, whose
versatility stems from the dependence of the potential on the
angular orientation between the rf field and the static trapping
field \cite{Lesanovsky2006}. In particular, one can create very
robust double-well potentials, which allow the coherent splitting
of a trapped Bose condensate \cite{Schumm2005}. In the experiments
presented here, we employ a setup similar to that in ref.
\cite{Hofferberth2006} where the combination of two rf fields
generated by auxiliary wires on the atom chip allows the
realization of a compensated double-well potential in the vertical
plane. For this double-well orientation, the observed interference
fringes in the atomic density are horizontal, parallel to the atom
chip surface. This allows us to image the interference pattern
along the transverse direction of the system (Figure
\ref{Fig:setup}).

\subsection{Preparing two 1d systems with a variable tunnel coupling}
To introduce a variable tunnel coupling between the two 1d
systems, we adjust the potential barrier between 3 and 8 kHz by
changing the amplitude of the rf fields (Figure
\ref{Fig:tunneling}). The actual barrier height is determined from
spectroscopic measurements of the rf potentials to a precision $<
1$ kHz \cite{Hofferberth2007}. To evaluate whether the two samples
in the resulting double-well potential can be considered as
one-dimensional, we compare the chemical potential $\mu$ and the
thermal energy $k_\text{B} T$ of the trapped ensembles to the
single-particle level spacing in the double-well. In the case of
large potential barriers, the transverse confinement of each
individual well can be approximated by a harmonic oscillator, and
the level spacing is given by the oscillator frequency
$\nu_\perp$. For small barriers, this approximation fails and one
has to numerically calculate the single particle states in the
transverse double-well potential. For the configurations shown in
figure \ref{Fig:tunneling} (b-d), we find energy separations
between the ground-state doublet and first excited states of
$\Delta = 2.8, 3.4,3.8\,$kHz, respectively. Consequently,
$\mu,k_\text{B}T \leq h \Delta$ for all configurations, justifying
the 1d treatment of the individual systems.

\subsection{Measuring the interference pattern and extracting the local relative phase}
We observe the interference pattern created by the two expanding,
overlapping atomic clouds using standard absorption imaging.
Employing diffraction-limited optics, optimized light path and
pulse duration, and a weak magnetic field to define a quantization
axis, we achieve atom shot noise limited imaging with 3 $\mu m$
resolution and a noise floor of $\sim 3$ atoms per 3x3 $\mu m$
pixel. We extract the local relative phase $\theta(z)$ by fitting
a cosine function with a Gaussian envelope to the observed density
distribution in each individual vertical pixel slice. The free
parameters of these fits are the relative phase $\theta$, the
contrast, and the fringe spacing. The width, amplitude, and center
position of the envelope are determined independently from a
Gaussian fit to the total integrated density pattern of the
central area of each image. From the measured phases, the
coherence factor is evaluated (compare eq. \ref{coh_fac_def}).
This has to be contrasted to the methods used in ref. \cite{Hadzibabic2006,Gati2006b}, where the interference patterns are summed up and the contrast is analyzed.

\subsection{Evaluating the coherence decay constant}
Several aspects must be considered when evaluating the decoherence
constant $t_0$. First, the time $t=0$ when the tunnel coupling
vanishes and the two condensates start to evolve independently,
changes for different atomic densities and trap parameters and has
to be evaluated for each data set separately. Higher atomic
densities \cite{Gerbier2004}, or stronger transverse confinement
lead to increased tunnel coupling as the overlap of the two matter
waves in the tunnelling region becomes larger \cite{Smerzi1997}.
The moment of the decoupling of the two systems can be estimated
by numerically calculating the time-dependent tunnel coupling
throughout the splitting. Experimentally, one observes a
qualitative change in the detected interference patterns
\cite{Roehrl1997,Schumm2005,Hofferberth2006} when the barrier
becomes higher than the chemical potential. In the case of the
widely split independent condensates, the uncertainty in the
determination of $t=0$ is much smaller than 1 ms.

The second aspect to consider is that for a finite system length
L, the coherence factor does not approach zero but converges to a
finite value. The average coherence factor calculated from a
limited number of samples is non-zero, even if these phases are
totally random. Such a limited number of samples for a system of
length L is given by the finite imaging resolution, and by the
finite phase-coherence length itself \cite{Petrov2000}, which in
our case is comparable to a single pixel width. This results in an
additional offset in the equilibrium coherence factor, which can
be determined from the number of data points used in the
calculation of $\Psi$ \cite{Fisher1993}.

\end{document}